\documentclass[%
 reprint,
 amsmath,amssymb,
 aps,
 superscriptaddress,
]{revtex4-2}

\usepackage{amssymb}
\usepackage{xcolor}
\usepackage{amsmath,braket,float,graphicx}

\begin{document}

\title{Influence of buffer gas on the formation of $N$-resonances in rubidium vapors}

\author{Armen Sargsyan}
\affiliation{Institute for Physical Research, National Academy of Sciences of Armenia, Ashtarak-2, 0203, Republic of Armenia}
\author{Rodolphe Momier}
\email[Corresponding author: ]{rodolphe.momier@u-bourgogne.fr}
\affiliation{Institute for Physical Research, National Academy of Sciences of Armenia, Ashtarak-2, 0203, Republic of Armenia}
\affiliation{Laboratoire Interdisciplinaire Carnot De Bourgogne, UMR CNRS 6303, Université Bourgogne Franche-Comté, 21000 Dijon, France}
\author{Claude Leroy}
\affiliation{Laboratoire Interdisciplinaire Carnot De Bourgogne, UMR CNRS 6303, Université Bourgogne Franche-Comté, 21000 Dijon, France}
\author{David Sarkisyan}
\affiliation{Institute for Physical Research, National Academy of Sciences of Armenia, Ashtarak-2, 0203, Republic of Armenia}

\date{\today}

\begin{abstract}
The $N$-resonance process is an accessible and effective method for obtaining narrow (down to subnatural linewidth), and contrasted resonances, using two continuous lasers and a Rb vapor cell. In this article, we investigate the impact of buffer gas partial pressure on the contrast and linewidth of $N$-resonances formed in the $D_1$ line of an $^{85}$Rb thermal vapor. $N$-resonances are compared to usual EIT resonances, and we highlight their advantages and disadvantages. Several rounds of measurements were performed with five vapor cells, each containing Rb and Ne buffer gas with different partial pressures (ranging from 0 to 400 Torr). This reveals the existence of an optimum Ne partial pressure that yields the best contrast, for which we provide a qualitative description. We then study the behavior of the $N$-resonance components when a transverse magnetic field is applied to the vapor cell. The frequency shift of each component is well described by theoretical calculations.

\end{abstract}



\maketitle

\section{Introduction}
\label{sec:intro}

\begin{figure}
\centering
\includegraphics[scale=1.8]{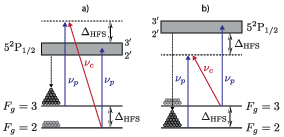}
\caption{Diagram of the $\Lambda$-system ($^{85}$Rb $D_1$ line) involved in the $N$-resonance process. The ground state level splitting is $\Delta_{\text{HFS}}~=~3036$~MHz. a) The probe laser is swept accross $F_g = 2 \rightarrow 2', 3'$ transitions while the coupling is fixed $\Delta_{\text{HFS}}$ higher than the $F_g = 2$ set of transitions. b) The probe laser is swept across $F_g = 3 \rightarrow 2', 3'$ transitions while the coupling is fixed $\Delta_{\text{HFS}}$ lower than the $F_g = 3$ set of transitions. In each case, the dotted arrow shows the population transfer due to optical pumping, which is reflected by the amount of grey beads on each of the ground states. The dotted line is a virtual level.}
\label{fig:level_diagram}
\end{figure}

$N$-resonances are narrow-band, all-optical Doppler-free absorptive resonances. They were first observed and studied by Zibrov et al. \cite{zibrov_observation_2002}. Initally referred to as \textit{three-photon absorption resonances} or \textit{three-photon EIT resonances} due to their resemblance with EIT resonances, they result from a two-field (probe and coupling) three-photon absorption process involving a two-photon Raman transition combined with optical pumping (from the probe beam). They are obtained in $\Lambda$-systems (as illustrated in Fig.~\ref{fig:level_diagram}). In this paper, we study the $D_1$ optical transition of $^{85}$Rb. The states involved in the process are thus $F_g = 2,3$ and the (unresolved) excited electronic state $5^2$P$_{1/2}$. The coupling beam is detuned from the excited state by the ground state hyperfine frequency $\Delta_{\text{HFS}}~\simeq~3.036$~GHz \cite{steck_rubidium_2019}. When the Raman condition $\nu_p - \nu_c = \Delta_{\text{HFS}}$ is fulfilled, a $N$-resonance appears in the absorption spectrum, in the form of an increase in absorption on top of the usual Doppler-broadened profile. Therefore, $N$-resonances differ from EIT resonances in that the latter result in a decrease in absorption \cite{wynands_precision_1999,fleischhauer_electromagnetically_2005,xu_optical-optical_2022,finkelstein_practical_2023}.
$N$-resonances are closely related to Electromagnetically Induced Absorption (EIA) resonances, which have been studied in vapors of other alkali atoms (Cs, Na and K) \cite{zibrov_observation_2002,lezama_electromagnetically_1999,sarkisyan_narrow_2009,bason_narrow_2009,hancox_lineshape_2008,sargsyan_n-type_2012,slavov_sub-natural_2014,hayashi_interference_2015,krasteva_observation_2020,sargsyan_formation_2022,brazhnikov_electromagnetically_2019}.
As in EIT, the linewidth of a $N$-resonance is strongly dependent on ground-state hyperfine decoherence.

When an external magnetic field $\bold B$ is applied, EIT and $N$-resonances behave similarly: the number of components in which they split depends on the hyperfine structure of the system (namely the total    angular momentum $\bold F$ of the lower ground states), the orientation of the magnetic field $\bold B$ with respect to the laser propagation direction $\bold k$ and the laser polarization. It was recently shown in \cite{mckelvy_application_2023} that EIT components formed in a magnetic field are useful tools that allow to retrieve both the direction and magnitude of the magnetic field. In principle, this could also be true for $N$-resonances. 

Previous studies \cite{zibrov_observation_2002,bason_narrow_2009,hancox_lineshape_2008,sargsyan_n-type_2012,slavov_sub-natural_2014,hayashi_interference_2015,krasteva_observation_2020,sargsyan_formation_2022} have shown that $N$-resonances exhibit enhanced contrast and reduced linewidth when a buffer gas, typically Neon, is introduced in the alkali vapor cell. However, in-depth investigations to pinpoint the optical buffer gas pressure for the formation of $N$-resonances have been lacking. Despite the extensive literature on the EIT process (see for example \cite{wynands_precision_1999,fleischhauer_electromagnetically_2005,xu_optical-optical_2022,finkelstein_practical_2023} and references therein), research on $N$-resonances remains limited. This highlights the importance of further investigation.
Due to their weak light shift compared to CPT resonances, $N$-resonances are promising candidates for small atomic frequency standards that can be realized with commercially available diode lasers \cite{slavov_sub-natural_2014, hayashi_interference_2015, novikova_cancellation_2006}.

In this work, we use five vapor cells with varying buffer gas pressures, spanning from 0 to 400 Torr. The objective is to investigate how these pressure conditions influence the contrast and linewidth of the $N$-resonance. By covering a broad spectrum of pressures, we aim to gain a comprehensive understanding of the behavior of the $N$-resonance under various experimental conditions. This allows us to observe how the spectral features behave and provides a valuable insight into the optimal buffer gas pressure conditions required for the formation of a narrow and contrasted $N$-resonance. Once the optimal pressure range is obtained, we examine the evolution of the resonance when an external magnetic field is applied and compare the frequency shift of its components with previous theoretical calculations.

\section{Experiment}
\label{exp}

\begin{figure}
\centering
\includegraphics[scale=1.5]{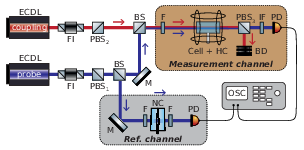}
\caption{Sketch of the experimental setup. ECDL - extended-cavity diode laser, FI - Faraday isolator, PBS - polarizing beam splitter, BS - beam splitter, F - neutral density filter, Cell + HC - 0.8 cm vapor cell containing a mixture of Rb and Ne (buffer gas) placed between Helmholtz coils, IF - interference filter, BD - beam dump, PD - photodetector, M - mirror, NC - nanometric-thin cell for the formation of reference spectra, OSC - oscilloscope.}
\label{fig:exp_scheme}
\end{figure}

The layout of the experimental setup is shown in Fig.~\ref{fig:exp_scheme}. Two VitaWave \cite{vassiliev_compact_2006} continuous external cavity narrow-band ($\delta \nu \sim 1$ MHz) diode lasers were used. Both lasers were tuned in the vicinity of the $D_1$ line of $^{85}$Rb. Two beams with mutually perpendicular polarizations were formed, hereafter referred to as probe ($\nu_p$) and coupling ($\nu_c$) beams, using polarizing beam splitters labelled PBS$_1$ and PBS$_2$. The probe laser frequency was tunable in order to scan the 5$^2$P$_{1/2}$ state, while the frequency of the coupling laser was fixed. The probe power was kept in the range 0.5 - 1 mW, while the coupling power could be increased up to 40 mW. Roughly 10$\%$ of the coupling laser was directed to a Dichroic Atomic Vapor Laser Locking (DAVLL) frequency locking scheme \cite{yashchuk_laser_2000}, omitted in Fig.~\ref{fig:exp_scheme} for the sake of clarity.

\begin{figure} 
\centering
\includegraphics[scale=1.5]{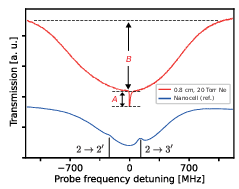}
\caption{Probe absorption spectrum (upper, red curve) scanning the $F_g = 2$ transitions of $^{85}$Rb $D_1$ line, recorded with a 0.8 cm cell containing a mixture of Rb vapor and Ne buffer gas ($P_{\text{Ne}} = 20$ Torr). The spectrum exhibits a $N$-resonance when the coupling laser is on, corresponding to the case depicted in Fig.~\ref{fig:level_diagram}a. The lower (blue) curve is a reference spectrum obtained with a nanometric-thin cell, as depicted in Fig.~\ref{fig:exp_scheme}.}
\label{fig:spectrum}
\end{figure}

The probe and coupling beams were combined with a beam splitter and directed to a 0.8 cm-long vapor cell containing Rb and Ne buffer gas. Several rounds of measurements were performed with cells having a different Ne partial pressure. In all cases, the temperature of the cell was 50~$^\circ$C, corresponding to a number density of the order of 10$^{12}~$cm$^{-3}$. The photodetector was then used to record the probe radiation passing through the cell, the coupling radiation being cut out by the analyzer labelled PBS$_3$. The cell was placed in the middle of three pairs of Helmholtz coils (Cell + HC), allowing the formation of a magnetic field if needed. 
A fraction of the coupling radiation (Ref. channel) was directed to a nanometric-thin cell with a thickness $L \sim \lambda = 795$ nm \cite{sargsyan_n-type_2012} to form a reference spectrum, as presented in Fig.~\ref{fig:spectrum} (lower, blue curve).

\begin{figure}
\centering
\includegraphics[scale=2]{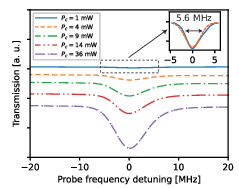}
\caption{Evolution of the $N$-resonance for different values of the coupling laser power $P_c$, recorded with a 0.8 cm cell filled with Rb and Ne ($P_{\text{Ne}} = 6$ Torr). The inset is a close-up on the $N$-resonance obtained with $P_c = 1$ mW, fitted with a Gaussian profile with a FWHM of 5.6 MHz.\label{fig:nres_power}}

\end{figure}

The red curve in Fig.~\ref{fig:spectrum} is an absorption spectrum of the probe beam $\nu_p$ passing through a 0.8 cm cell ($P_{\text{Ne}} = 20$ Torr) when the coupling beam $\nu_c$ is on. The spectrum contains a narrow $N$-resonance with a FWHM of 9 MHz and a contrast of around 20\%. Throughout this paper, the contrast is defined as the ratio of the depth of the $N$-resonance ($A$) to the probe-only absorption ($B$) \cite{kitching_chip-scale_2018}. The contrast increases with the power of the coupling laser, as can be seen in Fig.~\ref{fig:nres_power}, and can reach with this cell a maximum of 25\% (approximately 2.5 greater than the contrast of a $N$-resonance formed in a cell without buffer gas).  For a cell with 6 Torr Neon, the contrast can reach up to 40\%.

In Fig.~\ref{fig:level_diagram}, we present two diagrams of the energy levels of $^{85}$Rb $D_1$ line involved in the $N$-resonance formation process. Two configurations are presented, depending on whether the probe laser is scanning the $F_g = 2$ or $F_g = 3$ set of transitions. In diagram a) the probe field $\nu_p$, resonant with the transition between the lower-energy level of the ground state ($F_g = 2$) and the electronic excited state $5^2$P$_{1/2}$, pumps the atoms into the upper ground state $F_g = 3$  \cite{happer_optical_1972}, enhancing the probe transmission through the vapor, ultimately leading to population inversion. The coupling beam $v_c$ is detuned from the $F_g = 2$ transition. If the two-photon absorption condition $\nu_c - \nu_p = \Delta_{\text{HFS}}$ is fulfilled, atoms are driven coherently back to the lower energy level $F_g = 2$ via a two-photon absorption process \cite{sarkisyan_narrow_2009}, followed by a one-photon absorption from the probe bringing the atom to the excited state. The spectrum then shows a narrow, all-optical $N$-resonance induced by the three-photon non-linear process on top of a Doppler-broadened background caused by regular absorption, as it is shown in Fig.~\ref{fig:level_diagram}. The frequency of the $N$-resonance can be easily changed by tuning the frequency of the coupling field. In diagram b), a similar behavior is depicted. The probe field, resonant with the transition between $F_g = 3$ and the excited state, pumps the atoms into the lower-energy level of the ground state. In this case, if the condition $\nu_p - \nu_c = \Delta_{\text{HFS}}$ is fulfilled, atoms are driven coherently to $F_g = 3$, followed as before by a one-photon absorption from the probe bringing the atom to the excited state. More details about $N$-resonance formation are provided in section \ref{sec:discussion}.

Fig.~\ref{fig:nres_power} shows the dependence of the amplitude of the $N$-resonance ($P_{\text{Ne}} = 6$ Torr) on the coupling power. Five different curves are provided, which were respectively obtained for $P_c = 1,~4,~9,~14$ and $36$ mW (from top to bottom). The probe power was fixed at~$\sim$~1~mW in all cases. Since the diameter of both laser beams is $\sim$ 2 mm, the intensity $I$ for a laser of power $P$ [mW/cm$^2$] is $\sim 32~P$.

Note that reducing $P_c$ and $P_p$ to, say, 1 mW and 0.5 mW respectively allows to form a $N$-resonance with a smaller contrast (a few \%) but with subnatural linewidth, as shown in the inset of Fig.~\ref{fig:nres_power}. The resonance was fitted with a Gaussian profile (which is a good approximation for low coupling powers) and its FWHM is in that case around $5.6$ MHz. 

\begin{figure}
\centering
\includegraphics[scale=1.9]{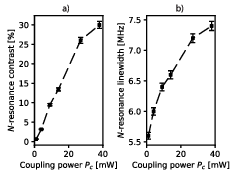}
\caption{Contrast (a) and Linewidth (b) of the $N$-resonance with respect to the coupling laser power $P_c$, each with $5\%$ error bars. The experimental parameters are the same as in Fig.~\ref{fig:nres_power}. The dotted lines are drawn to guide the eye.}
\label{fig:contrast_width}
\end{figure}

In Fig.~\ref{fig:contrast_width}a, we show the dependence of the $N$-resonance amplitude on the coupling power $P_c$. We observe an increase of the amplitude with the coupling power following a somewhat linear dependency (the dashed line are only to guide the eye). However, the increase in spectral width is much weaker with respect to the coupling power, as depicted in Fig.~\ref{fig:contrast_width}b. This is important when studying the splitting of the $N$-resonance in an external magnetic field to avoid overlapping of the different components.

\begin{figure}
\centering
\includegraphics[scale=1.9]{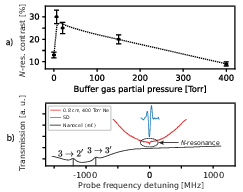}
\caption{a) Dependence of the contrast of the $N$-resonance on the buffer gas pressure measured in 0.8 cm long cells. Each black square (with error bar) is an experimental measurement performed in a separate cell with fixed buffer gas pressure. The dotted line is drawn to guide the eye. b) Probe transmission spectra obtained with 400 Torr Ne. The $N$-resonance is clearly visible on the red curve, and the second derivative (SD, blue curve) is shown to depict the increase in transmission.
\label{fig:contrast_pressure}}
\end{figure}

Fig.~\ref{fig:contrast_pressure}a shows the dependence of the contrast of the $N$-resonance on the Ne partial pressure, in the 0 to 400 Torr range. The coupling power is 1 mW and the probe power is 36 mW. It can be seen that the optimal Ne partial pressure is located between 6 and 30 Torr, where the contrast is maximum. It is interesting to note that this pressure is also optimal for the EIT process (see Fig.~8 in \cite{wynands_precision_1999}).

In Fig.~\ref{fig:contrast_pressure}b, we present a probe transmission spectrum recorded in a cell with 400 Torr Ne (red curve). The blue curve is the second derivative (SD) of the region highlighted by the dotted oval \cite{sargsyan_competing_2023}. In this case, the contrast is strongly reduced ($\simeq 10 \%$), and the SD allows to see the resonance better. The lower curve in Fig.~\ref{fig:contrast_pressure}b is a reference spectrum of the  $3 \rightarrow 2', 3'$ transitions of $^{85}$Rb.

It should be noted that despite the relatively low contrast of the $N$-resonance ($\sim 10$ \%) formed in a 400 Torr Ne vapor cell  shown in Fig.~\ref{fig:contrast_pressure}b, it is still easily recordable. The Saturated Absorption (SA) process, used in many experiments to form reference spectra, can not be seen in cells containing buffer gas (even as little as 0.1 Torr) \cite{demtroder_laser_2002}.

\begin{figure}
\centering
\includegraphics[scale=1.9]{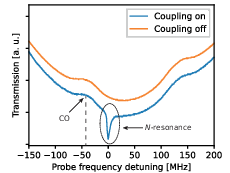}
\caption{Probe transmission spectra obtained with a $0.8$ cm cell containing pure Rb without buffer gas when the coupling beam is off (upper, orange curve) and on (lower, blue curve). The $N$-resonance is visible only when the coupling beam is on. The sub-Doppler structure is caused by the reflection of the laser beam on the inner surfaces of the cell windows. A cross-over (CO) resonance is visible, as formed in Saturated Absorption spectra \cite{zibrov_observation_2002}.\label{fig:onoff}}
\end{figure}

In Fig.~\ref{fig:onoff}, the upper orange curve shows the spectrum of only the probe radiation formed in a cell with pure Rb vapor,
while curve the lower blue curve shows the spectrum of probe radiation in the presence of the coupling radiation,
which contains a $N$-resonance with a linewidth of 6 MHz. Since the one-photon Doppler width of 
Rb $D_1$ line is $\simeq$~500~MHz \cite{thornton_velocity_2011,demtroder_laser_2002}, we observe strong narrowing of the $N$-resonance by $\sim$ 83
times.

\section{Behavior of the $N$-resonance in a magnetic field}

\begin{figure}
\centering
\includegraphics[scale=1.9]{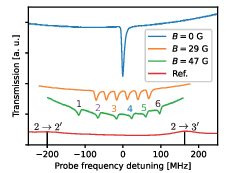}
\caption{Probe transmission spectra showing the behavior of the $N$-resonance when a transverse magnetic field ($\bold B \perp \bold k$) is applied. Due to the Zeeman effect, the single ($B = 0$) $N$-resonance is splitted into $6$ components when $B = 29$ G is applied. As expected, the components move away from each other as the magnetic field increases ($B = 47$ G). The $N$-resonance spectra were obtained with a 0.8 cm cell filled with Rb and Ne (20 Torr). The reference spectrum was obtained with a nanometric cell, as mentioned earlier.}.\label{fig:resonance_Bfield}

\end{figure}

\begin{figure}
\centering
\includegraphics[scale=1.9]{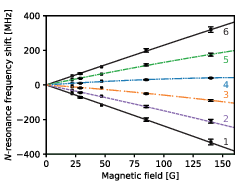}
\caption{Frequency shift of the $6$ components of the $N$-resonance (shown in Fig.~\ref{fig:resonance_Bfield}) as a function of the magnetic field. The lines were computed theoretically, the black squares are experimental measurements with $5\%$ error bars. Labelling and line style are consistent with Fig.~\ref{fig:lambda_systems}.
\label{fig:shift}}
\end{figure}

To study the evolution of the $N$-resonance in an external magnetic field, we placed a vapor cell (Rb + 20 Torr Ne) in the middle of a system of three pairs of Helmholtz coils. This allows to create a magnetic field in the desired direction while zeroing the laboratory magnetic field. 
The spectra presented  in Fig.~\ref{fig:resonance_Bfield} correspond to the case where a transverse magnetic field ($\bold B \perp \bold k$) was applied to the cell. The zero field resonance ($B = 0$, upper curve) splits into 6 equidistant components, all having the same linewidth of around 6~MHz. The components move away from each other as the magnetic field increases, as we see with the spectra corresponding to $B \simeq 29$ G and $B \simeq 47$ G. 
Under the influence of $B$, the ground levels $F_g = 2$ and $F_g = 3$ respectively split into 5 and 7 Zeeman sublevels. Since the frequency shifts of these levels are respectively $\mp \mu_B/3 \simeq  \mp 0.465$ MHz/G \cite{steck_rubidium_2019} (where $\mu_B$ is the Bohr magneton), the states are shifted in opposite directions and the distance between two adjacent components of the $N$-resonance can be easily estimated as

\begin{equation}
\Delta \nu = \frac{2\mu_BB}{3}\, ,
\end{equation}
yielding around 27 MHz for $B = 29$ G and 44 MHz for $B = 47$ G, which is consistent with what is seen in Fig.~\ref{fig:resonance_Bfield}. Equidistance of the components (as shown in Fig.~\ref{fig:shift}) is obtained in the small magnetic field regime ($B \ll B_0 = A_{\text{HFS}}/\mu_B$), where $A_{\text{HFS}}$ is the magnetic dipole interaction constant \cite{steck_rubidium_2019, olsen_optical_2011,wei_splitting_2005,mottola_electromagnetically_2023}. The value of $B_0$ for $^{85}$Rb is $\sim 0.7$ kG. For higher magnetic fields, i.e. when $\bold J$ and $\bold I$ are decoupled and $\ket{F,m_F}$ are not good quantum numbers, the Breit-Rabi formula can be used \cite{wynands_precision_1999,steck_rubidium_2019,wei_splitting_2005,mottola_electromagnetically_2023} to determine theoretically the frequency shifts of the states.

\begin{figure}
\centering
\includegraphics[scale=1.9]{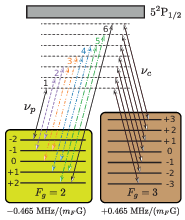}
\caption{Diagram of the $F_g = 2$ and $F_g = 3$ energy levels of $^{85}$Rb, which respectively split into 5 and 7 Zeeman sublevels when a magnetic field is applied. To form a $\Lambda$-system, the coupling and probe lasers must have the same upper state $\ket{F_e, m_{F_e}}$. Ten possible pairs of $\nu_c$ and $\nu_p$ frequencies are shown, the pairs for which the probe laser frequencies are the same are labelled $1$ to $6$. Components $1$ and $6$ only include a single probe transition.}\label{fig:lambda_systems}

\end{figure}

\begin{figure}[H]
\centering
\includegraphics[scale=1.9]{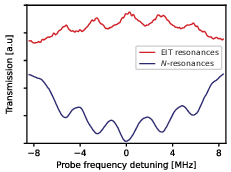}
\caption{Splitting of an EIT resonance (upper, red curve) and of a $N$-resonance (lower, blue curve) into five components in a longitudinal magnetic field ($B = 3$ G) with the following experimental parameters: $P_{\text{Ne}}  = 6$ Torr, $P_c = 1$ mW, $P_p = 0.2$ mW. Each component has a width of around 1.5 MHz. The distance between the lower and higher frequency components is 4~$\times$~0.93 MHz/G~$\simeq$~11~MHz.}\label{fig:EIT_comp}
\end{figure}

Fig.~\ref{fig:lambda_systems} is a diagram depicting the splitting of the $F_g = 2, 3$ states of $^{85}$Rb in a magnetic field. For the sake of clarity, we did not show the splitting of the $F_e = 2, 3$ levels of the excited state $5^2$P$_{1/2}$. In the diagram, ten possible pairs of probe/coupling frequencies which can lead to two-photon absorption from $F_g = 2$ with transfer of atoms to $F_g = 3$ are shown. It can be observed that the probe laser frequencies are in some cases the same (such groups of two transitions with the same frequencies are depicted by two arrows of the same color in Fig.~\ref{fig:lambda_systems}). These 6 pairs lead to the splitting of the $N$-resonance into 6 different components, as seen in Fig.~\ref{fig:resonance_Bfield} and \ref{fig:shift}. When $\bold B \parallel \bold k$, the $\sigma^+$ and $\sigma^-$ polarizations form 5 $\Lambda$-systems with the $F_g = 2$ and $F_g = 3$ levels of the ground state, and the $N$-resonance splits into 5 equidistant narrow-band components \cite{sargsyan_n-type_2012}. This configuration is shown in Fig.~\ref{fig:EIT_comp} for both EIT and $N$-resonances (a longitudinal magnetic field $B = 3$ G was applied). The linewidth of each component is about 1.5 MHz, 4 times narrower than the natural linewidth. However, it is clearly seen from the upper curve of Fig.~\ref{fig:EIT_comp} that the EIT signal is ``noisier'' than that of the $N$-resonance, which is explained by a higher contrast of the latter (i.e. higher signal-to-noise ratio).

\section{Discussion of the experimental results}
\label{sec:discussion}

Let us consider the diagram presented in Fig.~\ref{fig:level_diagram}b (all the following conclusions, with small modifications, are also true for Fig.~\ref{fig:level_diagram}a). As mentioned above, the purpose of adding a buffer gas (in this case, Ne) to a Rb cell is the following: 
\begin{enumerate}
\item A strong Rb-Ne collisional broadening of the $5^2$P$_{1/2}$ state occurs \cite{pitz_pressure_2014}, leading to greater absorption of the probe laser and, consequently, to an increase of the $N_2 - N_3$ value, where $N_2$ and $N_3$ are the populations of the ground states.
\item The speed of the atoms is greatly reduced, thereby increasing the time of flight of an atom through the laser beam, increasing the optical pumping process.
\end{enumerate}
A theoretical consideration of the $N$-resonance formation process is given in \cite{krasteva_observation_2020}. Here, we present a qualitative consideration of the process according to \cite{sarkisyan_narrow_2009}.

The amplitude $A$ of the $N$-resonance depends on the two-photon absorption resonance cross-section $\sigma_{\text{TPA}}$ such that
\begin{equation}
A \propto \exp [\sigma_{\text{TPA}}(N_2 - N_3)L], 
\end{equation}
where $L$ is the length of the vapor cell. The two-photon absorption resonance cross-section is defined as:
\begin{equation}
\sigma_{\text{TPA}} = \frac{\lambda^2}{16\pi^2}\frac{\Gamma_N}{\Gamma_{23}}\left(\frac{\Omega_C}{\Delta}\right)^2,
\label{eq:stpa}
\end{equation}
where $\Delta$ is the detuning of the coupling laser from the excited state $5^2$P$_{1/2}$, $\Gamma_N$ is the natural linewidth of the excited state ($\Gamma_N / 2\pi = 5.75$ MHz) and $\Gamma_{23}/2\pi = \gamma_{23}$ represents ground-state decoherence + other terms that may lead to line broadening \cite{wynands_precision_1999,fleischhauer_electromagnetically_2005}. The coupling Rabi frequency, $\Omega_C$, is defined as $dE_c/\hbar$, where $E_c$ is the coupling electric field and $d$ is the matrix element of the transition dipole moment of the $F_g = 3\rightarrow 5^2$P$_{1/2}$ transition. Estimates of the Rabi frequency can be obtained from \cite{krmpot_sub-doppler_2005}:
\begin{equation}
\frac{\Omega_C}{2\pi} \simeq \Gamma_N\sqrt{\frac{I}{8}},
\end{equation}
where $I$ is the laser intensity [mW/cm$^2$]. For $P_c = 30$ mW, we obtain $I = 960$ mW/cm$^2$ which yields $\Omega_c\simeq~62$~MHz.
Note that if the $N$-resonance is power-broadened, it is possible to estimate its linewidth with the formula given in \cite{finkelstein_practical_2023}: 
\begin{equation}
\gamma_{N_{\text{res}}} = \frac{\Omega_C^2}{\Gamma_\text{Dopp}}+\gamma_{23},
\label{eq:width}
\end{equation}
where $\Gamma_\text{Dopp}$ is the one-photon Doppler width \cite{demtroder_atoms_2010}:
\begin{equation}
\Gamma_{\text{Dopp}} = \sqrt{\frac{8k_B T \log(2)}{mc^2}}\, ,
\label{eq:dopp}
\end{equation}
with $\omega_0$ the $D_1$ transition frequency (here $\omega_0 = 377.107385690$ THz) and $m$ the atomic mass of $^{85}$Rb ($m = 1.409993199\times10^{-25}$ kg) \cite{steck_rubidium_2019,demtroder_laser_2002}. In our case, $\gamma_{23} \simeq 1$ MHz, which gives $\gamma_{N_{\text{res}}} \simeq 8.2$ MHz. From equation (\ref{eq:width}), it is clear that by reducing the coupling intensity, one can obtain a $N$-resonance with subnatural linewidth. An important fact to note is that amplitude depends on $\exp{(L)}$, therefore, $N$-resonances cannot be formed at small thicknesses, i.e. in micrometric-thin cells. However, since the dependence on $L$ is much weaker for EIT resonances, they can even be formed and detected in nanometric-thin cells \cite{tonoyan_formation_2023,sargsyan_comparison_2016}.
Another significant difference is that, when recording resonance fluorescence spectra during the EIT process, a narrow dip (decrease) can be observed on the fluorescence peak, while when a $N$-resonance is formed, we observe a narrow peak of increased fluorescence \cite{sarkisyan_narrow_2009}. 
Moreover, the increase of absorption exhibited in $N$-resonance spectra can be changed into a decrease of absorption in the presence of an additional laser field \cite{sarkisyan_conversion_2015}.

It is important to note that the two-photon absorption resonance cross-section $\sigma_{\text{TPA}}$ (eq. \ref{eq:stpa}) is inversely proportional to $\Delta^2$ ($\Delta$ is the detuning of coupling laser from the excited state $5^2$P$_{1/2}$). At high buffer gas partial pressures, a strong broadening of the excited state occurs \cite{pitz_pressure_2014}, resulting in an increase in $\Delta$. Since the buffer gas contributes to the population inversion required to form a $N$-resonance while simultaneously increasing the frequency detuning, there is an optimum for its partial pressure, as it is clearly seen in Fig.~\ref{fig:contrast_pressure}a.

\section{Conclusion}

In this article, we have studied the formation of all-optical high-contrast narrow-band $N$-resonance in $\Lambda$-systems of $^{85}$Rb $D_1$ line using two continuous-wave lasers. In the configurations studied in this paper, the $N$-resonance consists in an increase of absorption on top of the usual Doppler-broadened (+ additional broadening if the cell contains buffer gas) profile. We have performed, for the first time, the analysis in several Rb vapor cells containing different amounts of Ne buffer gas (from 0 to 400 Torr). It is demonstrated that the optimum of Ne partial pressure lies in the range 6 - 30 Torr. The $N$-resonances can be in many ways compared to usual EIT resonances:
\begin{enumerate}
\item $N$-resonances show an increase of both absorption and resonance fluorescence, while EIT resonances show a decrease of absorption and fluorescence.
\item To obtain narrow and contrasted $N$-resonances, the cell length must be quite large ($\ge$ $\mu$m), while EIT resonances can be formed in cells as thin as $100$ nm.
\item Under the same experimental parameters in the same $\Lambda$-systems, the contrast of the $N$-resonance is higher, meaning the signal-to-noise ratio is better (see Fig.~\ref{fig:EIT_comp}).
\end{enumerate}

In an external transverse magnetic field, six $N$-resonance components can be recorded (for $^{85}$Rb $D_1$ line), the frequency positions of which are very well described by theoretical calculations. This statement is also correction for EIT resonance. According to our expertise, forming a narrow and contrasted $N$-resonance in a centimetric vapor cell is easier than forming EIT resonances, while having a high signal-to-noise ratio is crucial for applications. This is explained by the strong dependence of the $N$-resonance amplitude on the intensity of the coupling laser shown in Fig.~\ref{fig:nres_power}. Thanks to narrow linewidth and high contrast, $N$-resonances can have a number of important applications (as much as EIT resonances) in a variety of fields, such as information storage, quantum communication, optical magnetometry or metrology \cite{wynands_precision_1999, fleischhauer_electromagnetically_2005,finkelstein_practical_2023}.

\section*{Acknowledgments}
This work was supported by the Science Committee of RA, in the frame of the project n°1-6/23-I/IPR, and was sponsored by the NATO Science for Peace and Security Programme under grant G5794.

\section*{Disclosures}
The authors declare no conflict of interest.



\bibliographystyle{elsarticle-num} 
\bibliography{refs.bib}


\end{document}